# Nano-fabrication and characterization of silicon meta-surfaces provided with Pancharatnam-Berry effect

PIETRO CAPALDO,[1,2,†] ALESSIA MEZZADRELLI,[2,3,†] ALESSANDRO POZZATO,[4] GIANLUCA RUFFATO,[2,3,‡] MICHELE MASSARI,[2,3] AND FILIPPO ROMANATO,[1,2,3,*]

[1]CNR-INFM TASC IOM National Laboratory, S.S. 14 Km 163.5, 34149 Basovizza, Trieste, Italy
[2]LANN, Laboratory for Nanofabrication of Nanodevices, EcamRicert, Corso Stati Uniti 4, 35127 Padova, Italy
[3]Department of Physics and Astronomy 'G. Galilei', University of Padova, via Marzolo 8, 35131 Padova, Italy
[4]ThunderNIL Srl, Via Foscolo 8, 35131 Padova, Italy
[†] Equally contributed
[‡]*gianluca.ruffato@unipd.it*
*filippo.romanato@unipd.it*

**Abstract:** In this paper the implementation of optical elements in the form of Pancharatnam-Berry optics is considered. With respect to 3D bulk and diffractive optics, acting on the dynamic phase of light, Pancharatnam-Berry optical elements transfer a phase which is geometric in nature by locally manipulating the polarization state of the incident beam. They can be realized as space-variant sub-wavelengths gratings that behave like inhomogeneous form-birefringent materials. We present a comprehensive work of simulation, realization, and optical characterization at the telecom wavelength of 1310 nm of the constitutive linear grating cell, whose fabrication has been finely tuned in order to get a π-phase delay and obtain a maximum in the diffraction efficiency. The optical design in the infrared region allows the use of silicon as candidate material due to its transparency. In order to demonstrate the possibility to assemble the single grating cells for generating more complex phase patterns, the implementation of two Pancharatnam-Berry optical elements is considered: a blazed grating and an optical vortices demultiplexer.

## 1. Introduction

Metasurfaces based on high-index dielectric materials have gained increasing attention due to their ability to locally manipulate the amplitude, phase and polarization of impinging light achieving high spatial resolution, low intrinsic losses and showing potential compatibility with standard industrial processes [1]. Unlike diffractive and refractive elements, the phase shaping is not introduced through optical path differences but results from the geometrical phase that occurs with the space-variant polarization manipulation in the spatial domain [2, 3]. The basic principle of metasurfaces optics is based on the generation of spatially-variant artificial birefringence by properly structuring the substrate surface. The form-birefringence can be induced by the nanopatterning of high resolution gratings in combination with the local control of the orientation of the grating ridges [4]. At a first glance, the metasurfaces are constituted of ultra-thin 1D dielectric grating pixels (Fig. 1) that have the periodicity in the regime of sub-wavelength respect to the light, and an arbitrary grating vector orientation relative to light polarization [5]. The local rotation of the gratings is proportional to the geometric phase, also known as Pancharatnam–Berry (PB) phase, whose local control allows the design of the phase map over the full area of the optics surface. The key-element of such a PBOE is varied pixel-by-pixel in order to transfer the desired geometric-phase pattern to the incident light, providing the space variety of grating orientation that can be viewed as continuously rotating wave plates with constant retardation and continuously space-varying fast axis.

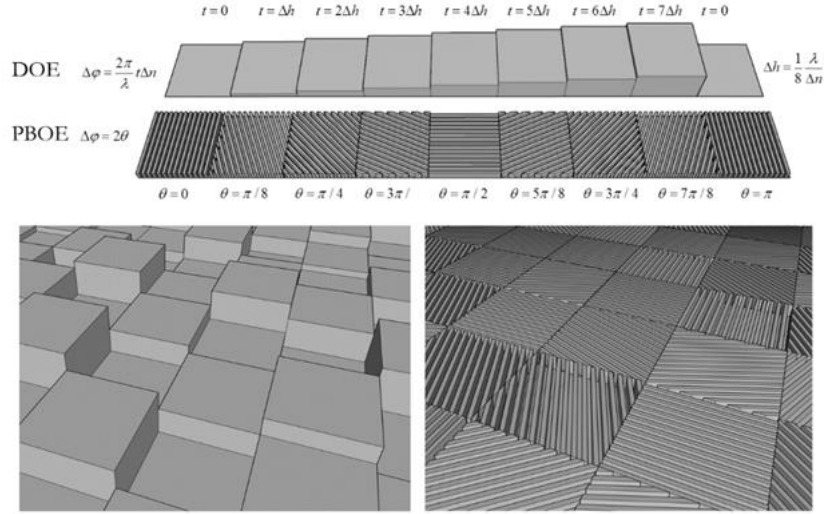

Fig. 1. Comparison between a diffractive optical element (DOE) on the left and a Pancharatnam-Berry optical element (PBOE) on the right, in the case of 8 phase levels. It is possible to infer the higher flatness and ease of fabrication of 2D with respect to 3D optics.

As a pioneer in this research field, Hasman and co-workers have first experimentally demonstrated the circularly polarized light conversion at a wavelength of 10.6 µm by using metallic gratings [6], and, later, using dielectric gratings [7-9] for beam shaping and splitting. Levy *et al.* have also used the rotated dielectric gratings to achieve polarization-dependent holograms at wavelengths of 1.55 and 10.6 µm [5, 10]. More recently, different optics designs have been engineered to control the wavefronts of incident electromagnetic waves and support beam steering and focusing functionality with high efficiency, enabling multipurpose optical devices which are inherently polarization-sensitive [11]. In particular, the ability of metasurfaces to scale traditionally bulky optical parts down to ultrathin components promises to enable ultraminiaturized optical systems [12, 13] that will increase the scope and the context of photonic applications showing, for instance, the capacity to manipulate the angular momentum of light in ways that exceed the capabilities of conventional optical components [14].

These concepts have been initially developed using plasmonic effects of metal gratings or nano-antennas served as model systems for wavefront engineering [15, 16], however the idea has been difficult to extend to high-efficiency devices operating at visible and near-infrared wavelengths due to absorption losses intrinsic to metal. Since then, complementary efforts have been made to extent the PB phase concept realizing metasurfaces made of dielectric materials [17], which can have low or negligible absorption losses depending on the material and operating wavelength. To date, a wide range of dielectric materials have been studied and implemented in metasurfaces at visible regime, for example titanium oxide and silicon nitride [18].

Among the many different combinations of optical design and materials, subwavelength silicon metasurfaces are reserving particular interest for near and mid-infrared applications [19]. Silicon is among the most commonly used materials due to its large dielectric constant and maturity of nanostructuring, and it has been implemented in metasurface design concepts based either on resonance tuning [20, 21] or effective medium subwavelength features [22,

23]. The operating wavelength range strongly depends on the crystallinity of silicon [24]: amorphous and polycrystalline silicon devices can efficiently operate in the infrared wavelength ranges, while single-crystal silicon devices can efficiently operate at near-infrared wavelengths due to their relatively low absorption losses.

For these reasons, a huge interest is going to be developed in telecommunications and silicon photonics for metasurfaces made of silicon working in the near infrared regime where silicon has real peculiar properties because of the high index of refraction and low absorption. Silicon PBOE can replace the conventional free-space bulky components for wavefront and polarization control, imaging, and even on-chip integration. The use of silicon will enable the integration of metasurface devices into existing technology platforms, due to the mainstream industrialization and mature development of silicon in electronics, on-chip photonics, and microelectromechanical systems [25].

In this view, this article will study the optimized conditions for the nanostructuring of PBOE made of crystalline silicon for near-infrared regime and specifically for telecom applications at 1310 nm wavelength, on which one of the two transmission windows of silica fibers are centered. Beside the main control parameter represented by the grating vector rotation, the efficiency conversion of the optical element depends on the geometric profile of the grating, i.e. on the duty-cycle and on its periodicity. Provided the period is much smaller than the wavelength in order to satisfy the regime of metasurface, it is necessary to determine the precise thickness as a function of the grating duty-cycle and period in order to achieve a precise control of the material optical parameters.

While a few examples of silicon sub-wavelength grating metasurfaces have been presented and tested in literature [5, 21, 22], to the best of our knowledge a complete experimental analysis of the fabrication processes and of the resulting optical response is still missing. The aim of this work is to bridge this gap, considering in depth the relation between the fabrication procedure of a subwavelength grating and the corresponding optical response, according to which a geometric phase delay is introduced. The thickness, period and duty-cycle of silicon grating for the determination of the optimal geometric phase has been computed and experimentally measured at a wavelength of 1310 nm and then computationally extended in the range 1260-1360 nm.

We performed a systematic control of the grating nanofabrication process and determined, by ellipsometric analysis and transmission measurements, the ordinary and extraordinary values of the effective refraction indices at the wavelength of 1310 nm. Electron Beam Lithography (EBL), Nano-Imprint Lithography (NIL) and Induced-Coupled Plasma Reactive Ion Etching have been exploited as nano-fabrication techniques to produce linear grating samples.

In order to demonstrate the possibility to assemble the single grating cells for the generation of more complex phase patterns, besides a simpler blazed grating, we also considered the realization for the first time of a PB mode-division demultiplexer for the sorting of optical beams carrying orbital angular momentum (OAM) of light [26]. Recently, the phase structuring of light has known an upsurge of interest in the so-called mode-division multiplexing, which exploits the spatial degree of freedom in order to increase the information capacity of optical links [27]. Modes carrying non-null OAM present a twisted wavefront induced by a phase term $exp(i\ell\varphi)$, being $\ell$ the amount of OAM per photon in units of $\hbar$. Modes with different $\ell$ values, i.e. different OAM, are orthogonal to each other and provide distinct channels for information transfer [28]. We considered the sorting method based on OAM-mode projection [29, 30] and we designed and fabricated a silicon metasurface performing both polarization-division multiplexing (PDM) and OAM-mode division multiplexing of 7 beams with OAM in the range from -3 to +3, for a total of 14 OAM channels.

## 2. PBOE design and simulations

### 2.1 Form-birefringence

Digital sub-wavelength gratings have birefringent properties arising from the particular geometric structure that leads to a form birefringence where the grating acquires uniaxial crystal properties [4]. The optical response is characterized by two different refractive indices related to the extraordinary and ordinary propagation axes, perpendicular and parallel to the grating lines, respectively. As it is shown in Fig. 2, if the electric field of an incident linearly-polarized beam is parallel to the grating vector, the interaction with the metamaterial is through the lowest refractive index ($n_{TM} = n_\perp$), while if the electric field is perpendicular to the grating vector, a higher refractive index ($n_{TE} = n_\parallel$) is experienced.

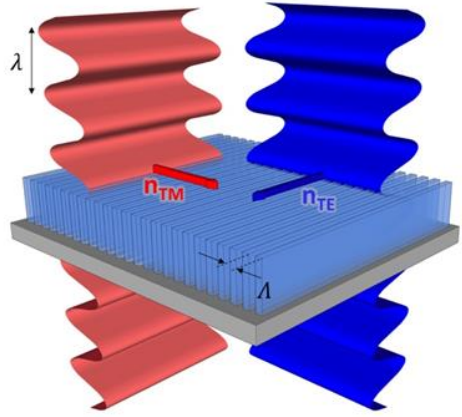

Fig. 2. TM polarization (in red) has the electric field parallel to the grating vector, whose modulus is given by $K=2\pi/\Lambda$. The orthogonal TE polarization (in blue) has the electric field parallel to the grating ridges. A qualitative comparison between wavelength and period of the grating is shown, which allows treating the grating as an effective form-birefringent material.

In general, a birefringent material with a thickness $d$ introduces a phase difference $\delta$ if crossed by a light beam, according to:

$$\delta = \frac{2\pi}{\lambda} d \left( n_\parallel - n_\perp \right) \tag{1}$$

It is convenient to describe the grating optical behavior with the use of Jones formalism [31] in $(x, y)$ basis representation, where the transmission function $T$ of the subwavelength grating at the position $(x, y)$ can be expressed as a rotated retarder by the matrix:

$$T(x,y) = R\left[\theta(x,y)\right] \cdot \tau(\delta) \cdot R\left[\theta(x,y)\right]^{-1} \tag{2}$$

$\tau(\delta)$ being the transmission matrix for a wave-plate with phase retardation $\delta$, $R$ the matrix for an optical rotator, and $\theta$ the angle with respect to the $x$-axis (representing the direction of a TM-polarized field). The two matrices have the explicit form:

$$R\left[\theta(x,y)\right] = \begin{vmatrix} \cos\theta(x,y) & -\sin\theta(x,y) \\ \sin\theta(x,y) & \cos\theta(x,y) \end{vmatrix} \tag{3}$$

$$\tau(\delta) = \begin{vmatrix} \exp(-i\delta/2) & 0 \\ 0 & \exp(+i\delta/2) \end{vmatrix} \tag{4}$$

In case of sub-wavelength grating, the matrix in Eq. (4) describes a form-birefringent medium whose phase retardation $\delta$ is determined by the grating profile in terms of period, duty-cycle and depth, and by the refractive index of the substrate material, and it is assumed to be constant for each pixel. The validity of last equation relies on the thin element approximation where the pixel size, $\Delta L$, the grating depth, $d$, and the wavelength, $\lambda$, fulfill the condition $\Delta L^2 >> d\lambda$.

For simplicity, we work with the helicity basis in which [1 $i$] and [1 $-i$] denote right-hand circular polarization and left-hand circular polarization, respectively (the normalization factor $1/\sqrt{2}$ has been omitted). In this representation, the resulting wave consists of two components: the zero-order, exhibiting the same polarization of the incident beam and no phase modification, and the diffracted order, exhibiting an orthogonal polarization and a phase term equal to twice the local orientation of grating, with a sign depending on the input handedness:

$$T\begin{pmatrix}1\\\pm i\end{pmatrix} = \cos\left(\frac{\delta}{2}\right)\begin{pmatrix}1\\\pm i\end{pmatrix} - i\sin\left(\frac{\delta}{2}\right)e^{\pm 2i\theta}\begin{pmatrix}1\\\mp i\end{pmatrix} \qquad (5)$$

It is worth noting that a maximum in diffraction efficiency can be achieved by assuming a retardation equal to $\pi$, i.e. the metasurface behaves as a rotated half-wave plate. In this case the zero-order term can be eliminated and the components of $T(x,y)$ assume the explicit form:

$$T\begin{pmatrix}1\\\pm i\end{pmatrix} = -ie^{\pm 2i\theta}\begin{pmatrix}1\\\mp i\end{pmatrix} \qquad (6)$$

Therefore, the desired phase modulation can be achieved just by varying the orientation of the sub-wavelength structure of each pixel, and a potentially continuous phase modulation can be obtained just by using a binary grating, eliminating the need of complicated multiple-step and grey-scale fabrication procedures (Fig. 1).

*2.1 Simulation and design*

In this work, the uniaxial form-birefringence properties of the basic PBOE cell (PBC) is obtained by structuring the medium with a sub-wavelength grating. It can be physically explained by the change of the boundary conditions experienced by the electromagnetic waves in the two directions, parallel and perpendicular to the grating vector. To express the two effective refractive indices as a function of the geometric features of the grating, two different approaches have been considered: Effective Medium Theory (EMT) and Rigorous Coupled-Wave Analysis (RCWA).

The Effective Medium Theory (EMT) is an approximate method where the sub-wavelength grating is described as a thin anisotropic birefringent meta-material with effective refractive indices [32, 33]. Under this approach, the effective refractive indices are formulated in terms of powers of the ratio between grating period and wavelength. Then, while in the zero-order EMT only the duty-cycle, i.e. the fraction between the grating and the surrounding medium, plays a role, in the second-order formulation the grating period starts to be a design parameter and the validity of the approximation is extended. On the other hand, the Rigorous Coupled-Wave Analysis (RCWA) is an exact approach and the effective refractive indices are given by numerical computations using a rigorous electromagnetic grating theory based on Fourier series expansions [34-36].

Simulations have been performed in order to obtain theoretical values for the effective refractive indices of the grating and to choose the best grating geometry in terms of thickness, period and duty-cycle, taking into account fabrication limits and experimental feasibility. Both EMT and RCWA simulations have been considered and implemented with specific custom MATLAB® codes. In all simulation sessions, an incidence normal to the sample interface has been considered.

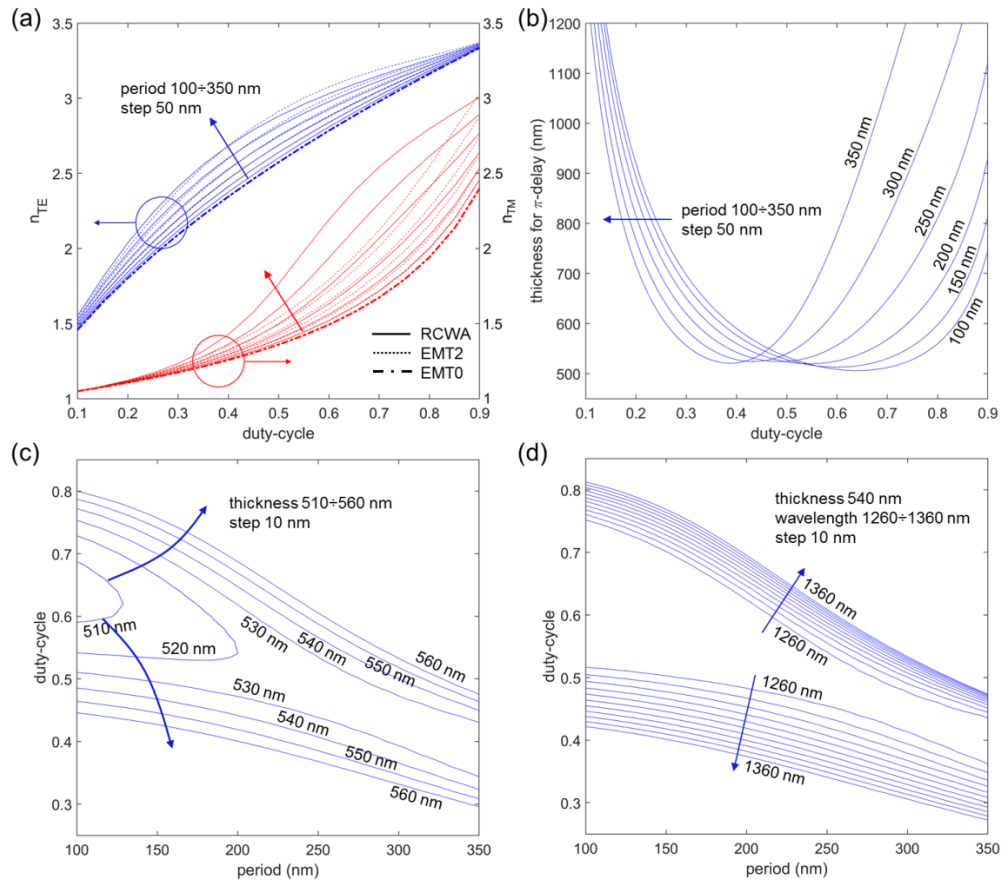

Fig. 3. (a) Comparison of effective indices for TE (blue) and TM (red) polarizations calculated with different numerical methods: zero-order effective medium theory (EMT0), second-order effective medium theory (EMT2), rigorous coupled-wave analysis (RCWA). Incident wavelength 1310 nm and different grating periods from 100 to 350 nm, step 50 nm, were considered at normal incidence. (b) Grating thickness providing π-delay between TE and TM polarizations as a function of the duty-cycle for different grating periods. (c) Optimal configurations of duty-cycle and period providing π-retardation for varying grating thickness from 510 to 560 nm, step 10 nm. (d) Optimal configurations of duty-cycle and period providing π-retardation for fixed grating thickness 540 nm, input wavelength in the telecom O-band from 1260 to 1360 nm, step 10 nm.

In Figure 3(a), the effective refractive indices are shown both for TE and TM polarizations, calculated with EMT and RCWA, as a function of the grating duty-cycle for different periods spanning from 100 to 350 nm, with a step of 50 nm, at 1310 nm. As expected, while for lower period values EMT estimations are closed to RCWA results, the two approaches diverge significantly for increasing periods, especially in the case of TM polarization. These results confirm the necessity of considering a rigorous approach for the grating design, instead of an approximated effective medium theory.

In Fig. 3(b), the corresponding thickness providing π-delay is reported for the same grating periods as a function of the duty-cycle. For a given grating period, the grating thickness dependence on the duty-cycle exhibits a *U*-shape, increasing abruptly for a duty-cycle close to the extreme values, since the two effective indices become equal (see Fig. 3(a)). From an experimental point of view, it appears more convenient to fix the grating thickness and optimize the grating profile. As more clearly shown in Fig. 3(c), for a given grating thickness above a threshold, there exist two distinct branches of optimal configurations

(providing π-delay). In principle, any point of the two branches could be selected equally. On the other hand, it is worth considering the behavior of such curves as a function of the input wavelength. In Fig. 3(d), the optimal-configuration branches are reported for a grating thickness of 540 nm and input wavelength spanning in the telecom O-band (1260-1360 nm). With respect to the lower branch, the upper branch exhibits a better tolerance to wavelength variation, which can be important for the design of broadband optical elements operating in the telecom infrared.

Several fabrication restrictions have been considered in order to obtain the best possible features. Line depth-to-width ratio (aspect ratio) should be kept as small as possible. Another limitation regards the duty-cycle: small values of it would prescribe very narrow lines that could collapse on each other. On the other side, a large duty-cycle will lead to wider lines separated by a small portion of free space, which could lead to merging problems.

In agreement to what previously explained, a period of 180 nm and a duty-cycle around 0.5 were chosen (on the lower optimization branch, see Fig. 3(c)). According to these values, it is possible to estimate the two effective refractive indices of grating using the RCWA analysis treated before: $n^{RCWA}_{TE}$=2.667, $n^{RCWA}_{TM}$=1.419. The depth of the grating, calculated by using Eq. (1) with a π-phase difference, is estimated around 530 nm.

An important limiting detail of RCWA is that it does not provide an analytical expression for effective refractive indices that, instead, can be obtained approximately by EMT theory with a lower computational effort. According to a second-order EMT, the ordinary and extraordinary refractive indices ($n^{EMT2}_{TE}$=2.667 and $n^{EMT2}_{TM}$=1.409) are close to the previous more rigorous estimations. Considering a higher period value around 290 nm, while keeping the duty-cycle equal to 0.5 (second optimization branch), the two methods clearly diverge: $n^{RCWA}_{TE}$=2.799 and $n^{RCWA}_{TM}$=1.574, $n^{EMT2}_{TE}$=2.811 and $n^{EMT2}_{TM}$=1.485. The optimal grating depth is around 535 nm.

## 3. Fabrication and optical characterization

### 3.1 Fabrication methods

Electron beam lithography is the ideal technique to transfer the computational patterns from a digital stored format to an imaging layer with high-resolution profiles [37, 38].

A JBX-6300FS JEOL EBL machine, 12 MHz, 5 nm lithographic resolution, working at 100 kV with a current of 100 pA was used. A thin layer of AR-P6200.09 resist (Allresist GmbH) was spun at 6000 rpm obtaining a thickness of about 170 nm and then baked in oven at 150°C for 30 minutes. An exposition dose of 225 μC/cm² was chosen from a previous dose matrix analysis.

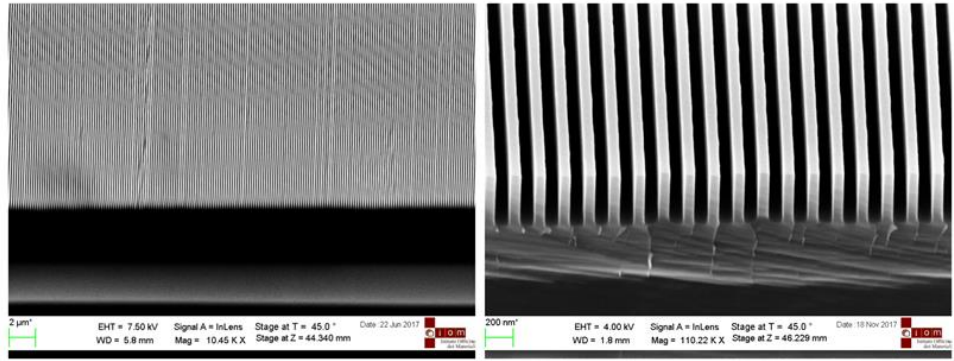

Fig. 4. SEM images of a sample at the beginning and at the end of the fabrication process. On the left, it is possible to appreciate the homogeneity of the grating on a wide scale. On the right, detail of the final sample at the end of the pattern transfer process.

For the fabrication of linear sub-wavelength grating with high aspect ratio a stamp process has been considered. The original EBL pattern was transformed in a imprinting mold for subsequent imprinting replica and ICP-RIE etching to achieve the final sample.

The resist pattern was transferred in the silicon substrate by means of STS MESC MULTIPLEX Reactive Ion Etching (RIE) plasma etching in Inductively Coupled Plasma-Reactive Ion Etching (ICP-RIE) configuration working at 13.56 MHz frequency. The ICP-RIE recipe used to transfer the resist pattern on the silicon substrate is composed by three sub-steps: $O_2$ stripping for the residual layer removal characterized by 40 sccm $O_2$ flow rate, 200 W of RF power of coil for the generation of plasma, 4 mTorr of chamber pressure, RF platen power of 10 W for the acceleration of ions to the sample. An etching step is performed in a three-gases-mix plasma ($C_4F_8$ at 60 sccm of flow rate, SF6 at 30 sccm and Ar at 10 sccm) with a coil RF power of 400 W, a RF platen power of 20 W (corresponding to an acceleration bias of 90 V) and a chamber pressure of 8 mTorr. Finally, an $O_2$ cleaning step was conducted characterized by 50 sccm of $O_2$ flow rate, a coil RF power of 800 W, RF platen power of 20 W and a chamber pressure of 20 mTorr.

The Thermal-NanoImprint Lithography process for the master fabrication was conducted using a Paul-Otto Weber hydraulic press with heating/cooling plates [39, 40]. The pattern generated by EBL is transformed into master stamp after a silanization process with Trichloro(1H,1H,2H,2H-perfluorooctyl)silane PFOTS [41, 42]. MR-I 7010E was spun on a silicon wafer at 2000 rpm to obtain a thickness of about 124 nm and a soft bake was conducted for 2 minutes at 140°C on the hot plate.

The T-NIL process is carried out at 80°C for an imprinting time of 20 minutes at a pressure of 100 bar, followed by cooling down at room temperature.

For the fabrication of the final samples (Fig. 4), the imprinting process was performed on a film of MR-I 7020E spun on silicon substrate at 1750 rpm to obtain a thickness of about 223 nm and baked for 2 minutes at 140 °C on the hot plate. The T-NIL process was conducted at 80°C at a pressure of 100 bar for 10 minutes followed by cooling down. Finally, the ICP-RIE etching was performed using the previous reported recipe.

*3.2 Ellipsometric analysis*

The analysis of ellipsometric data requires a model for the optical behavior of the material, defined in terms of its optical constants and layer thickness. In our case of interest, a thin film of uniaxial birefringent material over a bulk silicon substrate has been considered. Ordinary and extraordinary refractive indices have been assumed to follow a Cauchy model, and the initial theoretical values for the fit have been obtained from second-order EMT simulations.

A VASE Ellipsometer (J. A. Woollam) has been used in RAE (Rotating Analyzer Ellipsometer) configuration. The setup consists of a Xenon (Xe) lamp working in the range 190 nm - 2 μm with a monochromator and a focusing system. The polarization state is controlled by a first polarizer that can be set at the desired angle. The light beam, reflected by the sample into an elliptical polarization state, travels through a continuously rotating polarizer, and it is finally collected by a photodiode system.

For ellipsometric measurements, samples have been fabricated on single polished silicon wafers, in order to avoid the back-side reflection contribution which badly affects the fitting operation.

The data collection has been extended in an experimental range of wavelengths 1000 nm< $\lambda$< 1600 nm explored with a 5-nm step, while two different angles of incidence were chosen: 50° and 60°. Ellipsometric analysis output provides, besides the grating thickness, also the Cauchy coefficients of the biaxial index of refraction, $n_{TM}$ and $n_{TE}$, and eventually provides the phase delay of the samples, according to Eq. (1).

In Fig. 5, the calculated phase delay is shown as function of the grating depth. As expected, the phase delay scales linearly with the grating thickness. In table 1, data concerning nearly-optimized samples are reported.

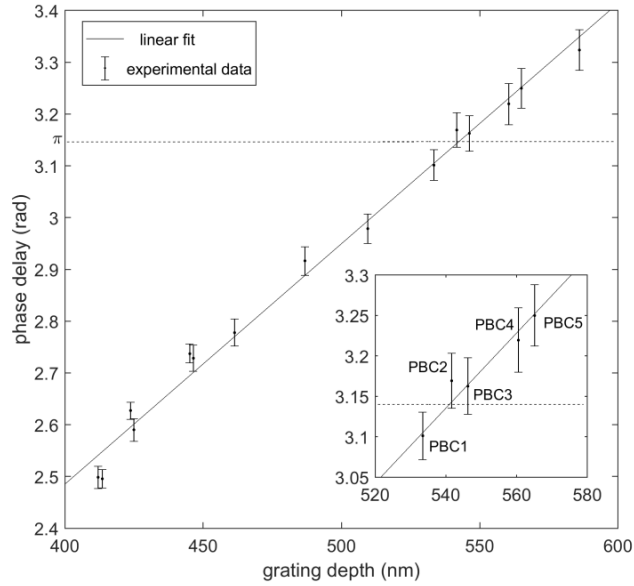

Fig. 5. Experimental data of the introduced phase delay as a function of the grating depth. Inset graph: zoom around nearly-optimized Pancharatnam-Berry samples (data shown in table 1).

### *3.3 Transmission analysis*

The phase-delay was measured also in transmission configuration. In this case the samples have been fabricated on double-polished silicon wafers in order to eliminate the back-surface scattering which dramatically reduces the amount of transmitted light. The original optical configuration of ellipsometry was modified as schematized in Fig. 6 in order to determine the two different components of the refractive indices. The incoming radiation from the ellipsometer source is converted into a circularly-polarized beam by a quarter-wave plate properly oriented at an angle of 45 degrees between the incident polarization and the fast/slow axes. A rotating polarizer analyzes the polarization state of the radiation transmitted by the PBOE.

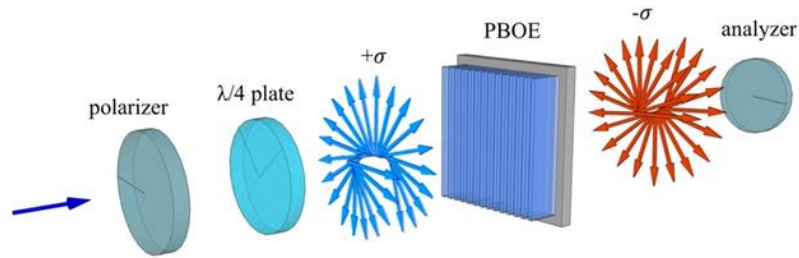

Fig. 6. Experimental setup employed to analyze the state of polarization exiting the sample in the transmission analysis measurements.

The experimental system can be studied using Jones matrix formalism according to which the sample can be viewed as a linear phase retarder with unknown retardation $\delta$:

$$E_{out} = \begin{pmatrix} \cos^2\theta & \sin\theta\cos\theta \\ \sin\theta\cos\theta & \sin^2\theta \end{pmatrix} \begin{pmatrix} e^{i\delta} & 0 \\ 0 & 1 \end{pmatrix} \begin{pmatrix} \cos\alpha & \sin\alpha \\ -\sin\alpha & \cos\alpha \end{pmatrix} \begin{pmatrix} i & 0 \\ 0 & 1 \end{pmatrix} \begin{pmatrix} \cos\alpha & -\sin\alpha \\ \sin\alpha & \cos\alpha \end{pmatrix} \begin{pmatrix} 1 \\ 0 \end{pmatrix} \quad (7)$$

where $\alpha=45°+\Delta$ is the angle between the input polarization axis and the quarter-wave plate slow axis, being $\Delta$ a systematic deviation which should be considered.

The equation for the intensity profile at the detector output as a function of analyzer angle $\theta$, dependent on the phase delay $\delta$ of the sample and on the residual misalignment angle $\Delta$, is used as fitting model for the experimental data:

$$I = a\left[1 - \cos(2\theta + 2b)\sin^2(2\Delta) - \cos(2\Delta)\sin(2\Delta)\sin(2\theta + 2b)\cos\delta + \sin\delta\right] \quad (8)$$

where $a$ and $b$ are fitting parameters controlling the vertical and lateral shift of the curve. By aligning the analyzer and the quarter-wave plate, it is possible to perform measurements of transmission analysis and hence extract the phase delay for each sample. The typical deviation $\Delta$ has been estimated to be $(1.216 \pm 0.136)°$.

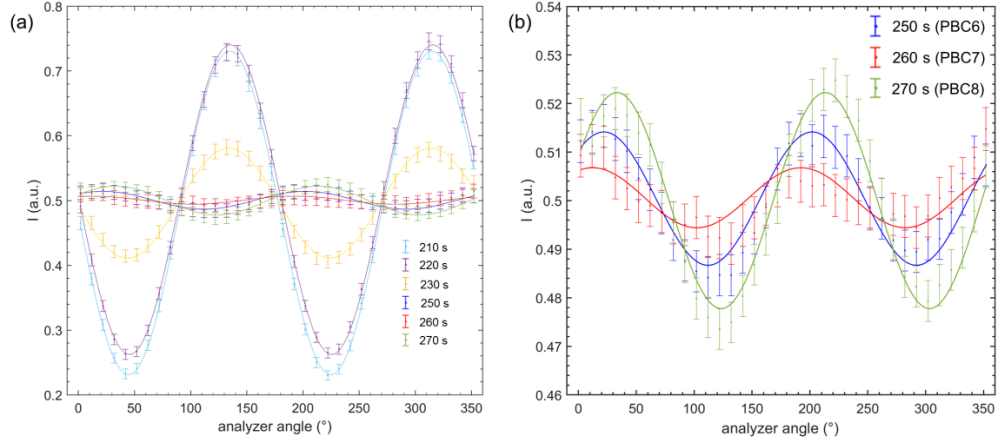

Fig. 7. (a) Experimental transmission dependence as a function of the analyzer angle for samples fabricated with different etching times, i.e. different grating thickness, fitted by means of Eq. (8). (b) Zoom for the nearly-optimized samples.

Table 1. Phase delay of Pancharatnam-Berry cell (PBC#) samples characterized via ellipsometry (PBC1-PBC5) or transmission analysis (PBC6-PBC8). Comparison between experimental results ($\delta_{exp}$) and numerical estimations ($\delta_{th}$) calculated with RCWA, assuming the tabulated values of period, duty-cycle and depth.

| Sample | period (nm) | duty-cycle | etching time (s) | depth (nm) | $\delta_{th}$ | $\delta_{exp}$ |
|---|---|---|---|---|---|---|
| PBC1 | 180 | 0.47 ± 0.01 | 250 | 533 ± 1 | 3.142±0.028 | 3.101±0.029 |
| PBC2 | 180 | 0.46 ± 0.01 | 260 | 542 ± 1 | 3.174±0.032 | 3.169±0.034 |
| PBC3 | 180 | 0.46 ± 0.01 | 260 | 546 ± 1 | 3.197±0.032 | 3.163±0.035 |
| PBC4 | 180 | 0.46 ± 0.01 | 270 | 560 ± 1 | 3.279±0.033 | 3.219±0.040 |
| PBC5 | 180 | 0.46 ± 0.01 | 270 | 565 ± 1 | 3.309±0.033 | 3.250±0.038 |
| PBC6 | 180 | 0.46 ± 0.01 | 250 | 540 ± 1 | 3.162±0.032 | 3.172±0.004 |
| PBC7 | 180 | 0.44 ± 0.02 | 260 | 546 ± 1 | 3.146±0.074 | 3.139±0.003 |
| PBC8 | 180 | 0.44 ± 0.01 | 270 | 554 ± 1 | 3.195±0.038 | 3.211±0.003 |

In Fig. 7, different samples of Pancharatnam-Berry cells (PBC) have been characterized via transmission analysis. The samples presenting a fairly constant intensity profile are those with a phase delay close to $\pi$, since the input circular polarization results to be completely converted into its opposite one. Table 1 reports the experimental phase-delay values, which are compatible with the expected numerical values calculated with RCWA.

## 4. Pancharatnam-Berry blazed grating

In order to demonstrate the possibility of implementing a diffractive optical element into its corresponding Pancharatnam-Berry counterpart, at first a simple optical system like a sawtooth blazed grating has been considered. The idea is to discretize the diffractive element into a mesh of differently-oriented pixels constituted of linear subwavelength gratings.

An ideal blazed grating deflects the incident beam by adding a component $\gamma$ to the linear momentum in the perpendicular direction. Its phase function is given by:

$$\phi(x,y) = \gamma x \tag{9}$$

where $\gamma = 2\pi/L$, $L$ being the period of the pattern. For an input wavelength $\lambda$ with incidence normal to the optical element, a deviation is introduced in the $x$-direction according to the relation:

$$\varphi = \arcsin(\lambda/L) \tag{10}$$

The continuous sawtooth profile of the corresponding diffractive optics has been divided into six discrete levels, to which a precise phase has been assigned according to the schematic representation shown in Fig. 1.

The same fabrication steps considered above for linear sub-gratings fabrication have been performed. The master has been fabricated with EBL and the pattern is a 2.52 x 2.52 mm square composed by 840 x 840 pixels with lateral dimension of 3 μm, which corresponds to a total rotation period of 18 μm. After silanization, the double imprinting process has been implemented. In Fig. 8, SEM images are shown for a sample exiting the first imprinting step.

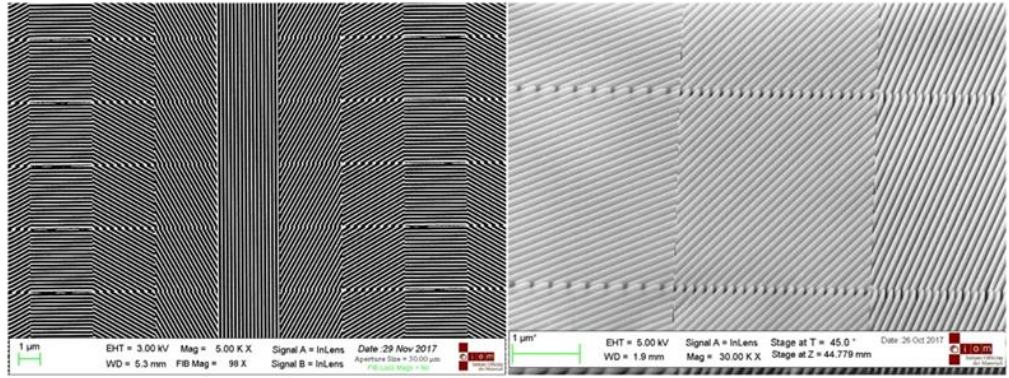

Fig. 8. SEM images of the sample labeled PBBG1. On the left, it is possible to appreciate a wide view of pattern homogeneity and its partition in pixel of different orientation. On the right, the attention is focused on single pixels.

To characterize these optical elements, a scatterometry analysis on the ellipsometer VASE has been carried out. Two quarter wave-plates are mounted mutually orthogonal at the two sides of the grating along the optical path in order to convert the linear polarization in circular one and vice versa. A static analyzer is placed after the second quarter wave-plate and before the detector, scanning a range of incidence angles from -10° to 10° with a 0.05° step.

The direction of the cross-polarized diffracted order is deflected towards positive (negative) angles if a left-handed (right-handed) polarization illuminates the sample (Fig. 9). Cross talking between different polarizations has been checked and reported.

The scatterometry graphs for three samples fabricated with thickness close to the optimal one are shown in Fig. 9. As foreseen the position of the diffraction peaks is fixed by the geometry of the PBOE and determined by eq. (10). On the contrary, the grating thickness is crucial in defining the conversion efficiency. RCWA simulations prescribe an optimal thickness of 545 nm for the experimental duty-cycle around 0.44.

The presence of a residual central zero-order peak, obtained by the measuring the same polarization of the input beam, can be noticed for samples PBBG1 and PBBG3, which have grating thickness around 493 nm and 582 nm, respectively. Instead, for the sample PBBG2, where the thickness of 542 nm is almost optimized, most of the incoming intensity is deflected on the two diffraction orders and the zero-order is approximately null.

This preliminary result proves the possibility to convert a diffractive optical element into a corresponding Pancharatnam-Berry component, where the desired shaping of the wavefront is obtained by introducing a spatially-variant geometric phase instead of manipulating the dynamic phase of light.

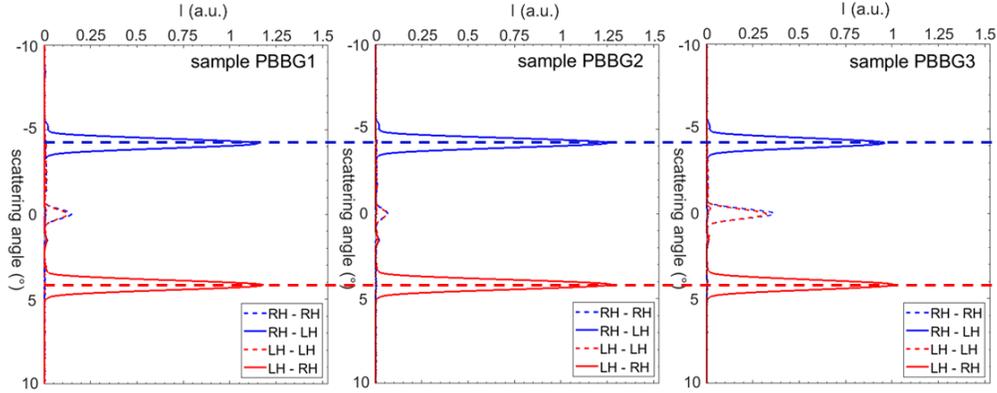

Fig. 9. Scatterometry graphs of samples PBBG1, PBBG2 and PBBG3 illuminated with both left-handed (LH) and right-handed (RH) circular polarization, measured for co-polarized (LH-LH, RH-RH) and cross-polarized output (LH-RH, RH-LH). The three samples have been fabricated with different depths: 493 nm (PBBG1), 542 nm (PBBG2), 582 nm (PBBG3).

## 5. Pancharatnam-Berry OAM sorter

### 5.1 Design

The phase pattern of a diffractive optics designed for analyzing the orbital angular momentum spectrum of the incident light field in the OAM set $\{\ell_j\}$ is given by the linear combination over a set of $n$ orthogonal modes $\{\psi_j = R_j(\rho,\vartheta)exp(i\ell_j\vartheta)\}$ as it follows [29]:

$$\Omega(u,v) = \arg\left\{\sum_{j=1}^{n} c_j R_j^* \exp\left[-i\ell_j\vartheta + i\alpha_j u + i\beta_j v\right]\right\} \quad (11)$$

where $\vartheta = arctan(v/u)$, $\{(\alpha_j, \beta_j)\}$ are the $n$ vectors of carrier spatial frequencies and $\{c_j\}$ are complex coefficients whose modulus is given arbitrarily, usually unitary, and their phases are free parameters of the task, fitted in such a manner that Eq. (11) becomes an exact equality [30]. The positions $\{(x_j, y_j)\}$ of the corresponding signal spots in far-field (Fig. 10) are given by:

$$(x_j, y_j) = \frac{f}{k}(\alpha_j, \beta_j) \quad (12)$$

$f$ being the focal length of the lens exploited for the far-field reconstruction in *f-f* configuration, and $k=2\pi/\lambda$. When an OAM-beam illuminates the optical element, its projection over the total OAM set appears in far-field, and a bright spot forms at the position corresponding to the carried OAM value (Fig. 10). When the optical element is realized into a metasurface form with total polarization conversion ($\pi$ retardation), the two orthogonal circular polarizations experience opposite phase patterns, according to:

$$\Omega^{(+)}(u,v) = \arg\left\{\sum_{j=1}^{n} c_j R_j^* \exp\left[-i\ell_j\vartheta + i\alpha_j u + i\beta_j v\right]\right\}$$
$$\Omega^{(-)}(u,v) = \arg\left\{\sum_{j=1}^{n} c_j R_j \exp\left[+i\ell_j\vartheta - i\alpha_j u - i\beta_j v\right]\right\}$$
(13)

and the corresponding signal spots are detected at the following distinct positions in far-field:

$$\beta^{(-)}(\ell_j) = -\beta^{(+)}(-\ell_j)$$
$$\alpha^{(-)}(\ell_j) = -\alpha^{(+)}(-\ell_j)$$
(14)

As a consequence of the polarization-dependent optical response, a beam carrying OAM equal to $\ell$ and left-handed circular polarization generates a bright spot in the far-field at a position which is symmetric to the spot produced by the right-handed circularly-polarized beam with opposite OAM.

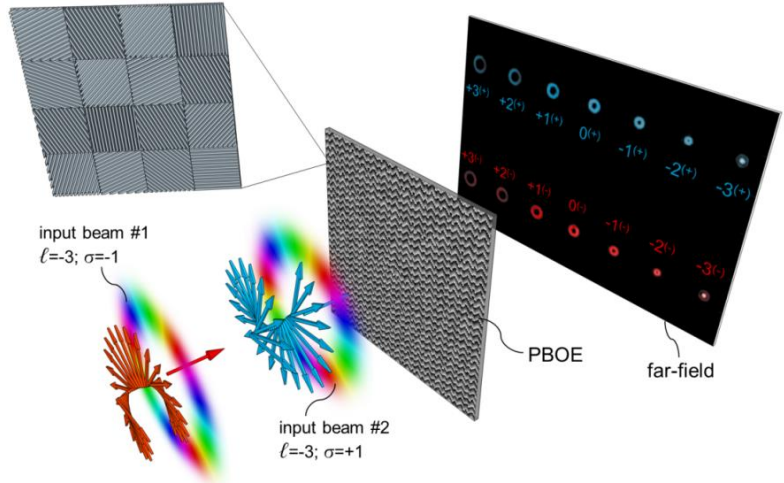

Fig. 10. Scheme of the Pancharatnam-Berry optics working principle for OAM-beam sorting with the method of optical beam projection. When a circularly-polarized OAM beam illuminates the optical element, a bright spot appears in the far-field, at a position depending on the carried OAM and on the polarization handedness.

A custom code implemented in MATLAB® has been used to calculate the phase pattern for the desired set of OAM values and the corresponding carrier spatial frequencies, taking into account definite limitations, as explained in [30], in particular the phase quantization into 16 equally-spaced levels. We limited the choice to OAM values in the set from -3 to +3 for a total of 7 OAM channels (Fig. 11(a)). The spatial frequencies were fixed in such a way that the far-field peaks are arranged over a line at equally spaced *x*-positions (see Fig. 11(b)). Considering Eq. (14) and the symmetry of the far-field channel constellation, it results that:

$$\beta^{(-)}(\ell_j) = -\beta^{(+)}(\ell_j)$$
$$\alpha^{(-)}(\ell_j) = \alpha^{(+)}(\ell_j)$$
(15)

As depicted in Fig.11(b), beams with the same OAM value and opposite handedness are detected at far-field positions which are symmetric with respect to the *y*-axis.

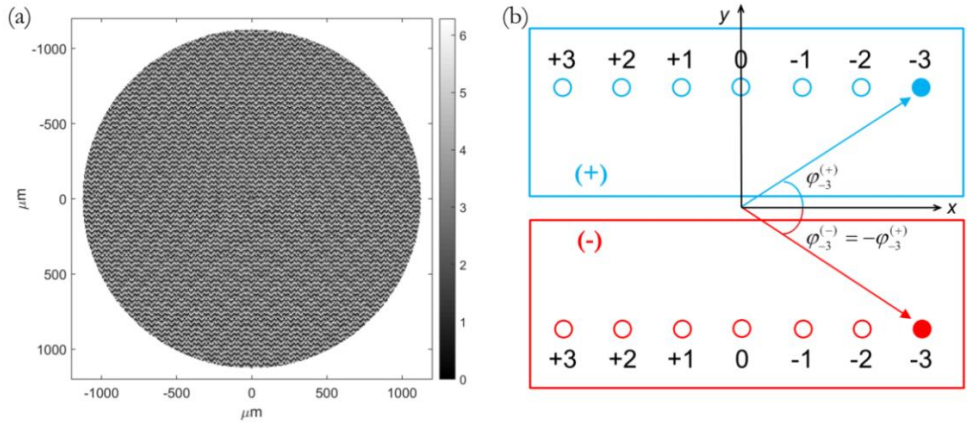

Fig. 11. (a) Numerical phase pattern for the sorting of OAM beams in the range from -3 to +3. 16 phase levels. Pixel size: 6.125 μm × 6.125 μm. Radius size: 256 pixels. (b) Far-field channel constellation for the given OAM set and circular polarization states.

With respect to the PB blazed grating, a different grating period of 290 nm has been considered, with duty-cycle around 0.5 nm, in order to select the upper branch group in the optimal configurations in Fig. 3(d). SEM inspections of the fabricated sample are shown in Fig. 12.

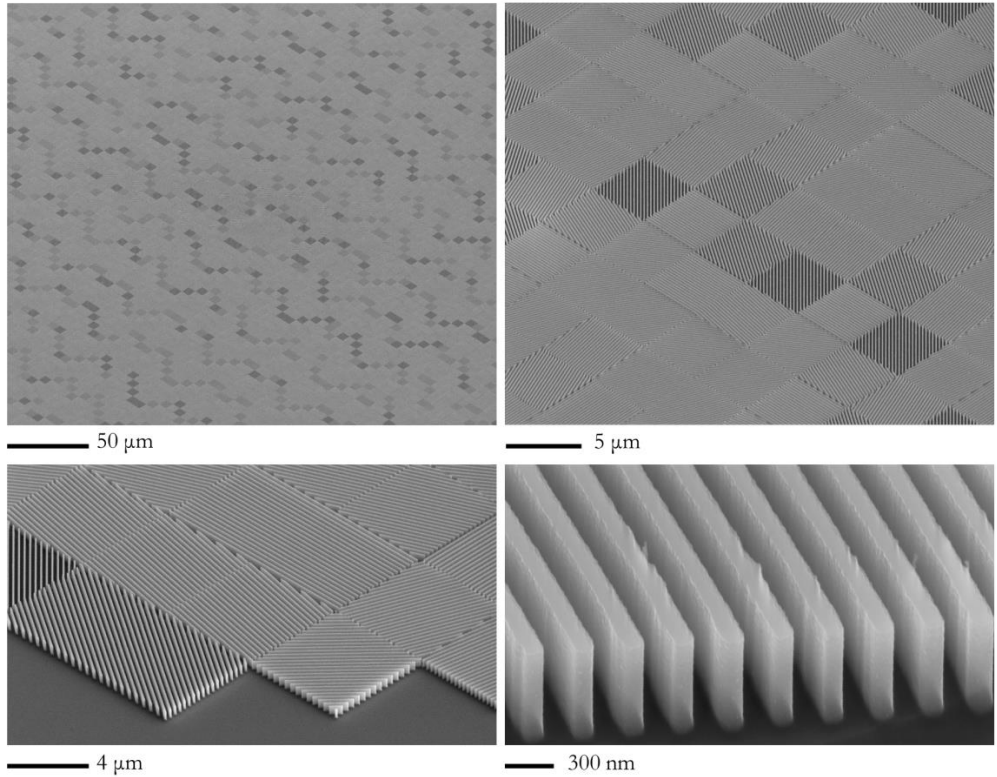

Fig. 12. SEM inspections of the fabricated PB demultiplexer on silicon substrate. Grating period Λ=290 nm, duty-cycle 0.5, thickness 535 nm, pixel size 6.125 μm. 16 rotation angles.

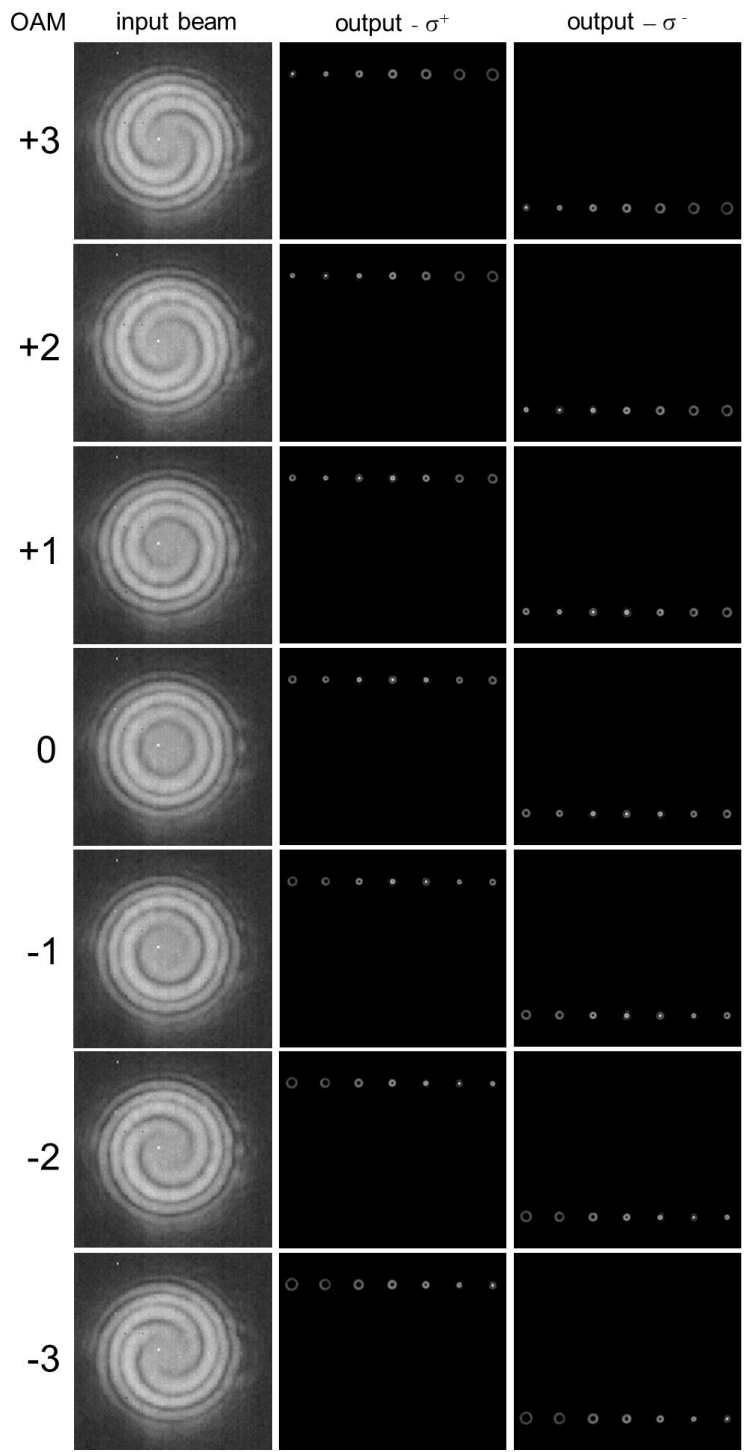

Fig. 13. Interference patterns of the OAM beams from -3 to +3 exploited for the sorter characterization. The number and handedness of the spirals denotes the carried OAM. Opposite-handedness far-fields for input right-handed and left-handed circular polarization states. As expected, the bright spot positions are in accordance with the theoretical channel constellation in Fig. 11(b).

*5.2 Optical characterization*

The optical behavior of the sample has been tested for illumination under beams carrying OAM, generated with a LCoS spatial light modulator (X13267-08, Hamamatsu, pixel pitch 12.5 μm). The output of a DFB laser ($\lambda$=1310 nm) is collimated at the end of the single mode fiber with an aspheric lens with focal length $f_F$=7.5 mm, linearly polarized and expanded before illuminating the display of the SLM. In order to generated an OAM beam with index $\ell$, the corresponding helical wavefront $exp(i\ell\varphi)$ is loaded on a doughnut intensity profile with a radius of 500 μm and a width around 200 μm and imaged on the SLM display using a phase and amplitude modulation technique [43]. Then, a 4-*f* system with an aperture in the Fourier plane isolates and images the first order encoded mode. A second 50:50 beam-splitter is used to split the beam and check the input beam profile with a first camera (WiSy SWIR 640U-S, pixel pitch 15 μm). Afterwards, the OAM beam illuminates the patterned zone of the silicon sample. The far-field is collected by a second camera (WiSy SWIR 640U-S) placed at the back-focal plane of a lens with focal length *f*=5.0 cm. A sequence of linear polarizer and quarter-wave plate is placed before and after the optical element, in reverse order, in order to generate and filter the desired circular polarization state. A Mach-Zehnder interferometric setup is added in order to analyze the phase pattern of the generated OAM modes. As Fig. 13 shows, the number and the helicity of the spiral arms in the interferograms denote the helical structure of the phase-front and the carried OAM value and sign.

The far-field has been collected for input optical beams with well-defined OAM in the set from -3 to +3 and circular polarization states. When an OAM beam illuminates the sorter, the far-field is the result of the projection of the field over the mode set for which the element has been designed. Therefore, a bright spot appears in correspondence of the detected OAM value, when it is present, at the position given by the channel constellation in Fig. 11(b). As Fig. 13 shows, the sorter separates the orthogonal polarization states onto two distinct lines, and the OAM beams are correctly detected according to the theoretical scheme.

## 6. Conclusions

In this work, we focus on the optimization of a linear sub-wavelength grating as constitutive basic unit for the design and implementation of high-efficient Pancharatnam-Berry Optical Elements in silicon working in the infra-red range. A three-step fabrication protocol, based on high resolution electron-beam lithography, thermal nano-imprinting lithography and ICP-RIE, has been optimized and finely tuned for the realization of sub-wavelength grating samples with a phase delay value closed to $\pi$ in order to obtain the maximum diffraction efficiency for an optimized single pixel. This effective half-wave plate represents the basic unit for the implementation of any phase pattern in the form of geometric phase optical elements.

Simulation performed with RCWA showed that the thickness for the definition of the $\pi$ phase delay also depends on the period and duty cycle of the grating. The experimental phase-delay determination confirmed the validity of the simulations that therefore can provide useful data for the design of PBOE.

The possibility to move from diffractive optical elements to PBOEs has been demonstrated with the design and test of two patterns: a blazed grating and a mode demultiplexer.

The sawtooth profile of a blazed grating has been discretized and each phase level has been converted into the corresponding rotation angle of the meta-pixel fast axis. A scatterometry analysis was used to confirm the goodness of our fabrication process for the realization of a quasi-ideal PB blazed grating. The optimal geometry of the PBOE obtained by matching the fabrication design specifications almost cancel the presence of spurious zero order peak contribution.

In a second step, we considered a more complex optical element performing the sorting of optical modes differing in their orbital angular momentum content. The optical characterization of the sample confirms the expected capability to sort circularly-polarized optical beams with different handedness and orbital angular momentum, providing a single optical element which can measure the total angular momentum of light.

This work paves the way for the realization of efficient Pancharatnam-Berry optical elements in the form of spatially-variant silicon sub-wavelength gratings for infrared applications, which could be exploited for the broad-band polarization-dependent phase-structuring and manipulation of light in the telecom field.

**Acknowledgments**

This work was supported by project New Optical Horizon region Lombardia, by project nanoMAX-nanoBRAIN of CNR, and finally by project Vortex 2 from CEPOLIPSE.